\documentclass[]{trbunofficial}
\usepackage{caption}
\usepackage{array}

\begin{document}

\title{Railroads as Cyber-Physical Systems: Cybersecurity Implications of Increasing Automation — Threats, Consequences, and Operational Resilience}

\TRBauthor{Dr. Syed Owais Athar}{School of Computing, University of Nebraska-Lincoln}{sathar2@huskers.unl.edu}[Lincoln, Nebraska, 68588][0000-0002-6253-0096]

\TRBauthor{Dr. Mohsin Kamal}{School of Computing, University of Nebraska-Lincoln}{mkamal3@unl.edu}[Lincoln, Nebraska, 87300][0000-0002-3893-4330]

\TRBauthor{Dr. Aemal Khattak}{Mid-America Transportation Center, Department of Civil \& Environmental Engineering,
University of Nebraska-Lincoln
}{khattak@unl.edu}[Lincoln, Nebraska, 68588][0000-0002-6695-9842]

\TRBauthor*{Dr. Muhammad Naveed Aman}{School of Computing, University of Nebraska-Lincoln}{naveed.aman@unl.edu}[Lincoln, Nebraska, 68588][0000-0002-4629-7589]

\AuthorHeaders{\textit{Athar, Kamal, Khattak, and Aman}}

\maketitle

\section{Abstract}

\hfill\break%
\noindent\textbf{Objectives:}~Rail automation is improving train protection, traffic management, inspection, maintenance, and operational visibility. It also increases dependence on software, communications, remote access, shared data, and external providers. This paper examines how automation changes the cybersecurity posture of railroad cyber-physical systems and identifies design measures that support operational resilience.

\hfill\break%
\noindent\textbf{Methods:}~The study uses a narrative review and threat-surface synthesis based on railroad automation literature, operational technology security guidance, Federal Railroad Administration and Transportation Security Administration materials, and documented rail cyber incidents. Automation functions are mapped to exposed assets, attack vectors, security properties, and operational consequences. The synthesis is organized around signalling, communications, onboard systems, remote-access pathways, and human oversight.

\hfill\break%
\noindent\textbf{Findings:}~Automation shifts rail risk from isolated electromechanical failure toward networked cyber-physical dependence. The most consequential threat surfaces involve train detection, location, signalling communications, grade crossings, remote diagnostics, vendor access, and artificial intelligence-enabled monitoring. Documented incidents show that even when fail-safe behavior prevents catastrophic outcomes, cyber events can create major operational disruption, supply-chain cascades, and national-security exposure.

\hfill\break%
\noindent\textbf{Novelty:}~The paper contributes a rail-specific taxonomy that links automation functions to cyberattack pathways, operational consequences, and resilience requirements. It frames cybersecurity as a transportation operations problem rather than a purely information-technology problem.

\hfill\break%
\noindent\textbf{Practical Applications:}~The findings support security-by-design procurement, information technology/operational technology segmentation, authenticated remote access, protected software updates, degraded-mode planning, trusted recovery, and meaningful human oversight. Rail operators and regulators can use the framework to evaluate automation deployments before cyber compromise becomes an operational disruption.

\newpage

\section{Introduction}\label{sec:intro}

Railroads are foundational transportation infrastructure. In the United States, freight rail spans almost 140,000 route miles and carries approximately 28 percent of domestic freight movement measured in ton-miles \cite{FRAFreight2025}. Rail service affects commodity flows, ports, energy distribution, manufacturing, emergency logistics, and passenger mobility. As operators pursue safety, capacity, asset visibility, energy efficiency, and cost control, rail functions are increasingly supported by automated protection, computerized dispatching, electronic interlocking, onboard diagnostics, wayside detectors, machine vision, condition monitoring, and data-driven maintenance \cite{LaitonBonadiez2022,Li2023,Kljaic2023}.

This transition is not a simple movement from manual to autonomous operation. Automation ranges from advisory systems to supervised task execution and control actions within prescribed operating boundaries. Positive Train Control (PTC), operating across all federally mandated U.S. route miles by the end of 2020, illustrates the operational value of automation because it enforces constraints intended to prevent train-to-train collisions, overspeed derailments, work-zone incursions, and movement through improperly positioned switches \cite{FRAPTC2023}. The same trend, however, changes the risk model: safe operation now depends on software logic, sensor integrity, wireless communication, positioning services, remote access, vendor systems, and shared data.

\subsection{Problem statement}

The same connectivity that enables automation weakens a separation that historically protected safety-critical railroad functions. Signalling, interlocking, and train-protection systems were once comparatively isolated from enterprise information technology (IT). Modern automated functions increasingly exchange data across onboard, wayside, control-center, maintenance, vendor, and enterprise environments. This IT/operational technology (OT) convergence expands operational capability, but it also creates attack paths and common dependencies through which one compromised service can affect train control, dispatching, inspection, and maintenance \cite{Guzman2019,Liagkou2021,NISTOT2023}.

The implication is a different assurance problem. Traditional fail-safe design manages random, independent, non-adversarial faults. Connected railway automation must also withstand deliberate actions such as unauthorized commands, manipulated sensor values, denial of service, credential compromise, and insecure software updates \cite{Collins2023,Thron2024}. Federal rail cybersecurity directives reflect that cybersecurity is now part of operational continuity, not only administrative IT compliance \cite{TSA2026}.

\subsection{Research questions and objectives}

This article asks four questions:
\begin{enumerate}[label=\textbf{RQ\arabic*.}]
  \item How do automation and IT/OT convergence alter the cyberattack surface of modern railway operations relative to legacy electromechanical control?
  \item Which threat surfaces carry the greatest potential for safety-critical and operationally cascading consequences?
  \item How can cyber-induced failures be distinguished from conventional technical faults, and what data or detection capabilities does this require?
  \item What combination of architectural, procedural, and human-oversight measures most effectively mitigates these risks without forgoing the benefits of automation?
\end{enumerate}
The objectives are to characterize attack-surface growth, map threat surfaces to physical and operational consequences, synthesize documented incident lessons, and identify resilience measures suitable for rail automation. Methodologically, the paper is a narrative review and threat-surface synthesis rather than a statistical incident-frequency study. It draws on peer-reviewed rail automation and cybersecurity literature, U.S. federal guidance, OT-security guidance, standards-oriented sources, and publicly documented incidents. The incident set is illustrative rather than exhaustive, selected to show how cyber compromise has reached rail operations through direct control-channel abuse, IT disruption, and supplier dependency.

\subsection{Scope and contributions}

The analysis emphasizes heavy mainline and freight rail, while drawing selective comparison to metro and communications-based train-control contexts where continuous train-to-wayside communication makes IT/OT coupling especially visible. Functionally, it spans signalling and communications, rolling-stock and onboard systems, dispatching, maintenance, remote access, and enterprise interfaces. The paper contributes: (i) a rail automation threat-surface taxonomy; (ii) an incident-based account of how cyber events become transportation consequences; and (iii) a resilience framework linking security architecture, degraded operation, trusted recovery, and meaningful human oversight.

The framework is intended for use before deployment as well as during operations. Rail agencies, host railroads, suppliers, and regulators can apply it during procurement, safety-case review, vendor-access approval, communications migration planning, incident-response exercises, and degraded-mode drills. In each setting, the question is the same: which automated function is being trusted, what evidence supports that trust, and how can the railroad continue safely if the supporting data or service is unavailable?

\section{Background: Rail Automation Systems}

Rail automation is a layered ecosystem of train-control, onboard, trackside, communication, and management functions. The cybersecurity question is therefore not whether a single device is secure, but whether the information flows that support automated decisions remain authentic, available, and operationally interpretable across the rail system.

\subsection{Grades of Automation (GoA 0–4)}

\begin{figure}
    \includegraphics[width=\linewidth]{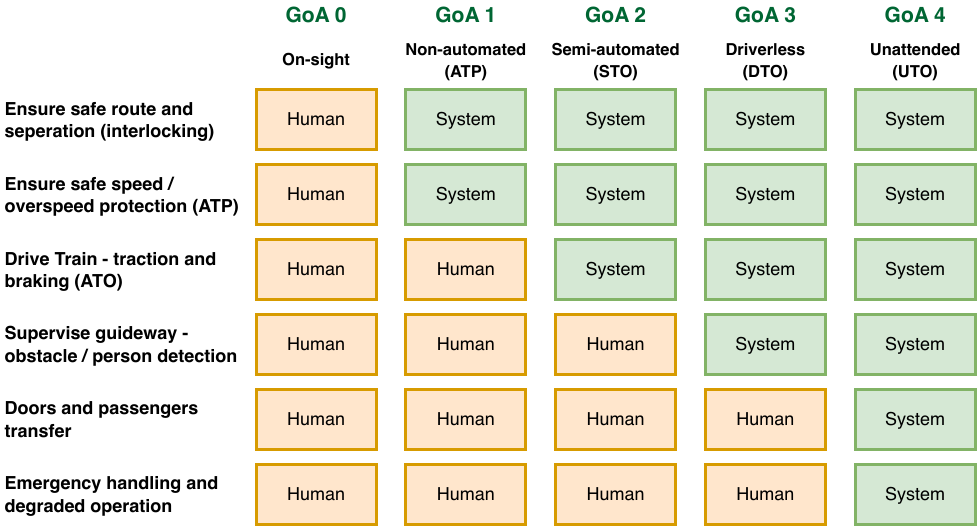}
    \caption{Grades of automation and the transfer of oversight to machines.}
    \label{fig:figure1}
\end{figure}

Grades of Automation (GoA) describe how operational responsibility shifts from personnel to automated systems. At \textbf{GoA 0}, the driver performs all functions; at \textbf{GoA 1}, Automatic Train Protection (ATP) supervises manual driving; at \textbf{GoA 2}, Automatic Train Operation (ATO) controls traction and braking while a driver supervises; at \textbf{GoA 3}, the system drives without a driver but with onboard staff; and at \textbf{GoA 4}, all driving, guideway supervision, passenger transfer, and emergency functions are automated. The cybersecurity relevance is direct: higher GoA levels concentrate safety-critical dependence on software, sensing, and communication whose integrity must be assured. Figure \ref{fig:figure1} summarizes this progressive transfer of oversight.

\subsection{Signalling and Train Control}

The safety and movement layer is commonly decomposed into ATP, ATO, and Automatic Train Supervision (ATS). ATP enforces speed and movement authority, ATO automates driving within the ATP envelope, and ATS regulates traffic above the individual train. European Train Control System (ETCS), Communications-Based Train Control (CBTC), and PTC implement these ideas in different operating environments. Their architectures differ, but they share a common cyber-physical property: movement decisions depend on the trustworthiness of onboard devices, trackside devices, radio links, control centers, and configuration data \cite{FRAPTC2023,Collins2023,Singh2021}.

\subsection{Interlocking Evolution: Relay → Electronic → Computer-Based}

Interlocking is the vital logic that permits only safe, non-conflicting route settings. As interlocking has moved from mechanical and relay-based arrangements toward electronic and computer-based implementations, safety logic has increasingly become software-defined. That shift improves monitoring and maintainability, but once interlocking is connected for diagnostics, updates, or remote management, cybersecurity becomes part of the safety argument.

\subsection{Rolling-Stock Control: TCMS and Ethernet Backbones}

Onboard control is coordinated through Train Control and Management Systems (TCMS), which integrate traction, braking, doors, diagnostics, and passenger-information functions. Ethernet backbones and onboard-to-ground diagnostic links improve data availability but also bring IT-style exposure onto rolling stock, especially when remote maintenance or software updates are enabled.

\subsection{Supporting Operational Technology}

Automation also depends on OT assets such as SCADA systems, traction-power controls, wayside controllers, balises, radio equipment, and field cabinets. These assets translate physical railroad conditions into digital state. If they are spoofed, delayed, misconfigured, or made unavailable, automated functions may execute correct logic using untrustworthy inputs \cite{Chen2015,FRAConnectedRail2020}.

\subsection{Communications Bearer: GSM-R and FRMCS}

Continuous automation requires a dependable communications bearer. GSM-R and emerging FRMCS/5G architectures support train-to-ground voice and data, while CBTC systems rely on continuous train-to-wayside radio. These bearers are safety-relevant because a loss of availability may stop trains, while a loss of authenticity or integrity may cause systems or personnel to trust false information.

\subsection{The IT/OT Convergence Thesis}

\begin{figure}
    \includegraphics[width=1\linewidth]{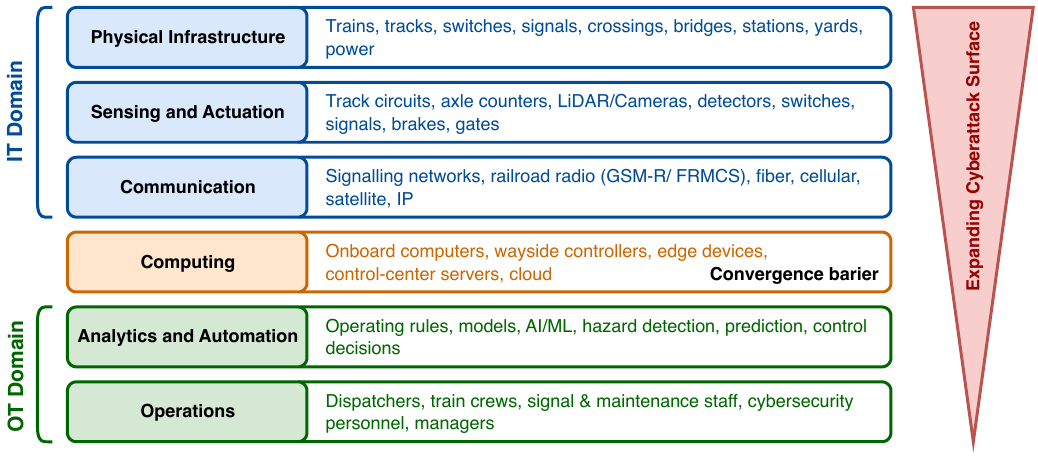}
    \caption{Six layers of a railway cyber-physical system, from trackside assets and sensors to communications, computing, automation, and day-to-day operations.}
    \label{fig:figure2}
\end{figure}

Modern railways can be viewed as six interacting layers: physical infrastructure, sensing and actuation, communication, computing, analytics and automation, and operations. Across those layers, modernization replaces localized indications with shared digital state. Figure \ref{fig:figure2} captures the resulting IT/OT convergence: the same integration that enables higher automation also expands the attack surface from limited and localized to distributed, interconnected, and dependent on shared services \cite{Guzman2019,Liagkou2021}.

\section{Cybersecurity Threat Landscape}

Integration across the cyber-physical stack reshapes rail security in four ways: it expands the attack surface, exposes railway operations to varied threat actors, creates connectivity-driven attack vectors, and introduces adversarial failure modes that traditional safety engineering was not designed to address.

\subsection{Attack-Surface Expansion Through Automation and Connectivity}

Automation adds components and trust relationships: field devices, onboard networks, radio links, control-center systems, maintenance laptops, vendor connections, software-update mechanisms, cloud platforms, and interfaces with external organizations. Industry 4.0 and OT-security literature identify connectivity, remote access, cloud integration, and IT/OT convergence as recurring sources of exposure \cite{EfeIsik2020,Liagkou2021,NISTOT2023}. Unlike enterprise IT, railway OT directly affects physical movement and must preserve safety, reliability, timing, and availability. The dominant security properties are therefore integrity, authenticity, and availability.

\subsection{Threat Actors and Objectives}

Cyber threats originate with actors whose objectives range from extortion to sabotage. Table \ref{tab:threat_actors} summarizes the actor classes most relevant to automated railway systems. The practical point is that cybersecurity planning must assume both financially motivated disruption and strategic attacks during geopolitical tension, while also accounting for insiders and compromised maintenance access \cite{NISTOT2023,TSA2026}.

\begin{table}[!ht]
  \caption{Threat actors targeting automated railway systems.}
  \label{tab:threat_actors}
    \renewcommand{\arraystretch}{1.2}
    \begin{tabular}{|
    >{\raggedright\arraybackslash}p{2.5cm} | 
    >{\raggedright\arraybackslash}p{3.5cm} | 
    >{\raggedright\arraybackslash}p{4.8cm} | 
    >{\raggedright\arraybackslash}p{4.0cm}|}
      \hline
      \textbf{Actor} &
      \textbf{Primary Objective} &
      \textbf{Characteristic Methods} &
      \textbf{Illustrative Rail Incident} \\
      \hline

      Nation-state/ state-nexus &
      Sabotage, espionage, and strategic disruption during geopolitical conflict &
      Stealthy long-term access, destructive/wiper malware, exploitation of public-facing and cloud vulnerabilities &
      Iran railways wiper attack (2021); Ukraine state railway cyberattack (2025) \\ \hline

      Organized crime/ ransomware &
      Financial extortion &
      Ransomware-as-a-service (RaaS), credential theft, encryption of IT and increasingly OT-adjacent systems &
      Pittsburgh Regional Transit ransomware incident (2024) \\ \hline

      Hacktivists &
      Political or ideological disruption and public visibility &
      DDoS attacks, website defacement, increasingly blended with ransomware; targeting aligned with geopolitical events &
      Belarusian railway attacks (2022); pro-Russia DDoS campaigns against European railways \\ \hline

      Insiders &
      Grievance, coercion, or inadvertent human error &
      Misuse of legitimate credentials, privileged maintenance access, and bypass of external security controls &
      Credential and maintenance-access misuse across railway systems \\ 

      \hline
    \end{tabular}
\end{table}

The actor categories overlap in practice. Ransomware may be used for financial extortion, political signaling, or disruption of military logistics. This convergence matters for rail because operational availability is itself a safety and public-service requirement.

\subsection{Attack Vectors}

Connectivity opens concrete pathways into safety-relevant systems. Table \ref{tab:attack_vectors} relates the principal vectors to their mechanisms, targets, and threatened security properties.

\begin{table}[!ht]
  \caption{Principal attack vectors in automated rail systems.}
  \label{tab:attack_vectors}
    \renewcommand{\arraystretch}{1.2}
    \begin{tabular}{|
    >{\raggedright\arraybackslash}p{2.8cm} |
    >{\raggedright\arraybackslash}p{4.3cm} |
    >{\raggedright\arraybackslash}p{4.6cm} |
    >{\raggedright\arraybackslash}p{3.0cm}|}
      \hline
      \textbf{Vector} &
      \textbf{Mechanism} &
      \textbf{Primary Target} &
      \textbf{Property at Risk} \\
      \hline

      Wireless links (train--wayside radio, positioning) &
      Spoofing, replay, jamming, and denial-of-service attacks &
      ETCS/CBTC communications, Radio Block Centre (RBC), and GNSS-based positioning &
      Integrity, Authenticity, Availability \\ \hline

      Remote diagnostics / maintenance access &
      Compromised remote sessions, maintenance laptops, and vendor VPNs &
      Train Control and Monitoring System (TCMS), health-monitoring platforms, and wayside controllers &
      Authenticity, Integrity \\ \hline

      Supply chain / vendor firmware and updates &
      Malicious or insecure software updates and third-party compromise &
      Onboard and wayside software, and shared vendor-managed services &
      Integrity, Authenticity \\ \hline

      Unpatched / unauthenticated legacy protocols &
      Exploitation of protocols lacking authentication or encryption &
      Legacy signalling systems and railway radio (e.g., unauthenticated GSM-R commands) &
      Authenticity, Integrity \\ \hline

      Physical trackside access &
      Tampering with field devices, cabinets, or transponders &
      Balises, trackside object controllers, and level-crossing equipment &
      Integrity, Availability \\

      \hline
    \end{tabular}
\end{table}

Two features distinguish these vectors from generic IT intrusion. First, several exploit trust relationships that automation actively creates, including vendor access and remote updates. Second, rail asset lifetimes keep legacy protocols in service long after their original threat assumptions expire.

\subsection{The Safety–Security Distinction}

\begin{figure}
    \includegraphics[width=\linewidth]{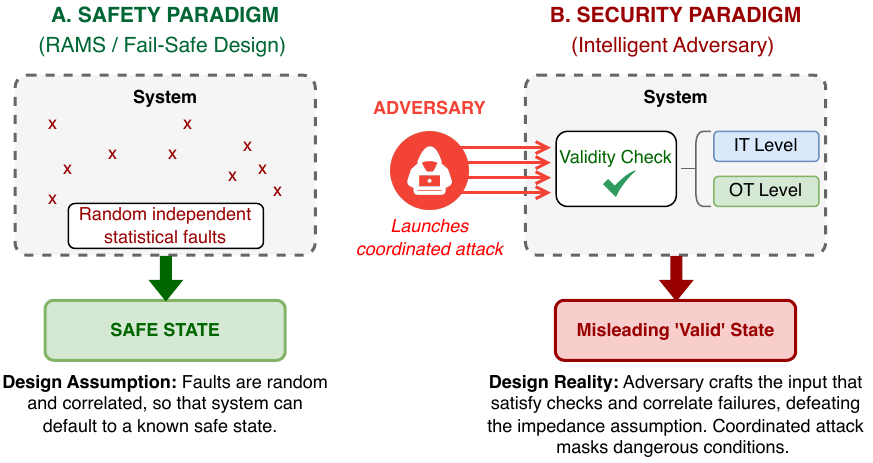}
    \caption{Random-fault safety model in contrast to intelligent-adversary security model.}
    \label{fig:figure3}
\end{figure}

Railway safety engineering assumes random, independent, statistically bounded failures. Fail-safe design then moves the system to a conservative state such as a stopped train or signal at danger. Cybersecurity adds an intelligent adversary who may craft plausible data, coordinate failures, replay legitimate messages, or exploit credentials. Figure \ref{fig:figure3} contrasts these paradigms. The distinction is operational: a cyber event may trigger safe stopping and still create network-wide denial of service, or it may create a superficially valid but unsafe state. Safety and cybersecurity therefore have to be argued together \cite{Collins2023,Chen2015,NISTOT2023}.

\section{Vulnerability Analysis}

\begin{figure}
    \includegraphics[width=1\linewidth]{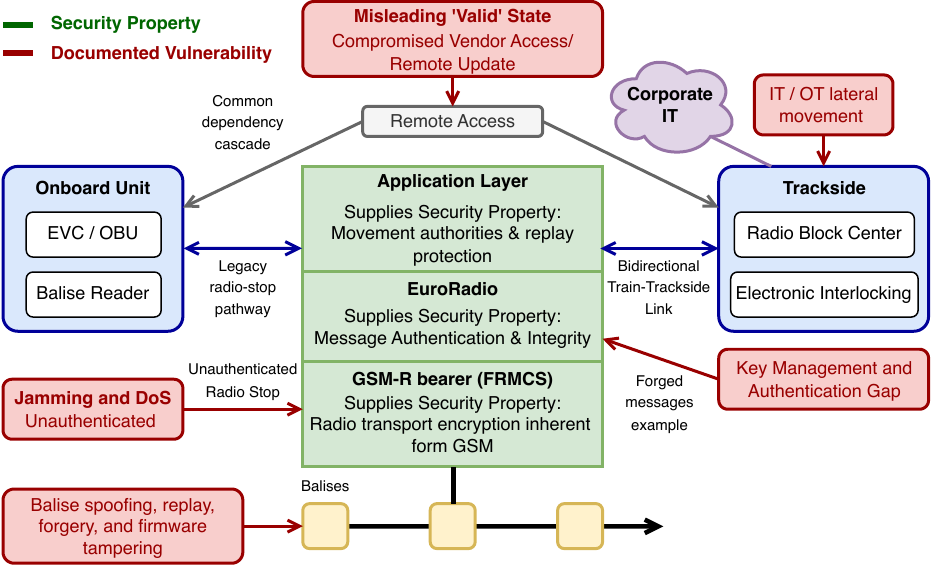}
    \caption{Attack surface of the ETCS/ERTMS train–trackside communication stack.}
    \label{fig:figure4}
\end{figure}

The preceding threat vectors become concrete at the level of protocols, radio bearers, trackside devices, interlocking interfaces, and vendor dependencies. Figure \ref{fig:figure4} locates several of these exposures on the ETCS/ERTMS communication stack.

\subsection{Legacy Protocols Lacking Authentication and Encryption}

Rail communications carry movement authority, location, train state, and emergency messages across systems with long asset lives. The 2023 Polish radio-stop incident illustrates the risk: attackers reportedly transmitted unauthenticated radio-stop commands that halted trains without compromising enterprise IT \cite{ReutersPoland2023}. The general lesson is that a legacy command channel can remain operationally powerful even when it lacks modern authentication, encryption, monitoring, or replay protection.

\subsection{Wireless Spoofing and Jamming Against CBTC and ETCS Bearers}

Continuous train-control paradigms depend on radio availability. Jamming may force a conservative stop, preserving immediate safety while reducing line capacity. Spoofing and replay are more dangerous when messages appear valid to automated systems. Where functions rely on Global Navigation Satellite System (GNSS) position or timing, spoofing and meaconing can corrupt the shared state used for control, dispatching, or crossing timing \cite{FragaLamas2017,FRAConnectedRail2020}.

\subsection{Balise and RBC Message Spoofing and Replay}

Balises, wayside controllers, and radio-control nodes are consequential because automated systems may treat their messages as authoritative. A forged, replayed, delayed, or tampered message can corrupt position, movement authority, or handover state. Physical exposure of trackside assets amplifies the risk because a cyber-physical attacker may combine local tampering with network manipulation.

\subsection{SCADA/PLC and Interlocking Exposure Once Bridged to Corporate IT}

SCADA, programmable logic controllers, and computer-based interlockings are increasingly bridged to enterprise systems for monitoring, diagnostics, and maintenance. Once bridged, an enterprise intrusion may become a pathway toward OT. The problem is aggravated by legacy industrial protocols, constrained devices, and remote administration channels that may not provide strong authentication or logging \cite{Aloseel2021,NISTOT2023}.

\subsection{Supply-Chain and Single-Vendor Dependency}

The final vulnerability is structural. Rail systems often depend on a small set of suppliers, long-lived proprietary platforms, remote vendor access, and shared update mechanisms. A compromised supplier, credential, software image, or common service can create correlated failure across functions that appear independent. This common-mode dependency directly challenges safety analyses that assume independent component failure \cite{Guzman2019,NISTOT2023}.

Table \ref{tab:vulnerability_classes} consolidates the five classes.

\begin{table}[!ht]
  \caption{Vulnerability classes in automated rail systems.}
  \label{tab:vulnerability_classes}
    \renewcommand{\arraystretch}{1.2}
    \begin{tabular}{|
    >{\raggedright\arraybackslash}p{2.8cm}| 
    >{\raggedright\arraybackslash}p{3.0cm}| 
    >{\raggedright\arraybackslash}p{3.5cm}| 
    >{\raggedright\arraybackslash}p{2.5cm}|
    >{\raggedright\arraybackslash}p{3.0cm}|}
      \hline
      \textbf{Vulnerability Class} &
      \textbf{Affected System / Protocol} &
      \textbf{Attack Technique} &
      \textbf{Property at Risk} &
      \textbf{Documented Evidence} \\
      \hline

      Unauthenticated legacy radio &
      Analogue railway radio (radio-stop); GSM-R (inherited weaknesses) &
      Spoofed stop command; eavesdropping &
      Authenticity, Availability &
      Poland railway incident (2023) \\ \hline

      Wireless bearer exposure &
      GSM-R/FRMCS (ETCS), CBTC radio, GNSS positioning &
      Jamming, denial of service, spoofing, and meaconing &
      Availability, Integrity, Authenticity &
      ERTMS risk analyses; GNSS spoofing research \\ \hline

      Balise / RBC message manipulation &
      Eurobalise, Radio Block Centre (RBC), EuroRadio &
      Spoofing, replay, message forgery, and firmware tampering &
      Integrity, Authenticity &
      Birmingham EuroRadio analysis; ERTMS attack-tree studies \\ \hline

      IT/OT bridge exposure &
      SCADA/PLC, computer-based interlocking &
      Lateral movement; insecure ICS protocols (Modbus/S7Comm); Ethernet sniffing and spoofing &
      Integrity, Availability &
      Aman (2026), Section~4.3; rail ICS protocol research \\ \hline

      Supply-chain / single-vendor &
      Vendor firmware/software; update and remote-access channels &
      Malicious or insecure software updates; common-mode compromise &
      Integrity, Authenticity, Availability &
      Aman (2026), Section~4.3; ERTMS risk analysis \\

      \hline
    \end{tabular}
\end{table}

\section{Case Studies: Documented Incidents}

\begin{figure}
    \includegraphics[width=\linewidth]{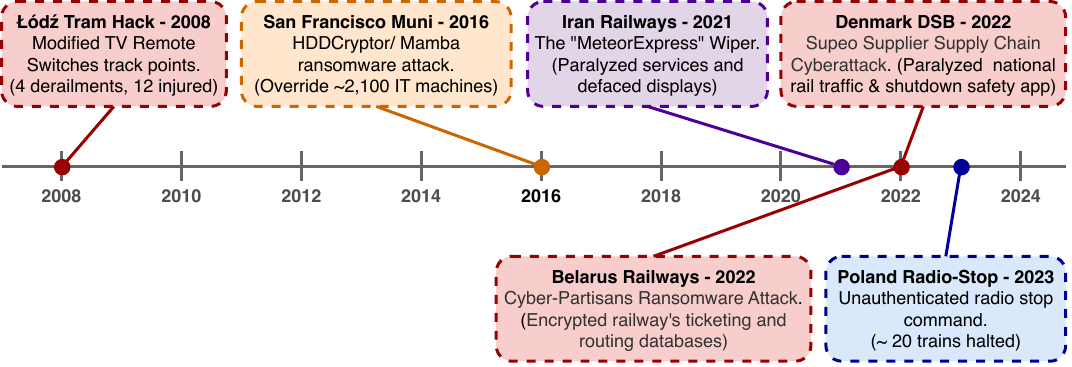}
    \caption{Timeline of selected railway cyber incidents from 2008 to 2023.}
    \label{fig:figure5}
\end{figure}

Documented incidents show that rail cyber risk reaches operations through several pathways: direct manipulation of a control channel, ransomware or wiper malware affecting peripheral IT, and supplier-service failure. Figure \ref{fig:figure5} places representative incidents on a timeline; Table \ref{tab:rail_incidents} summarizes the operational lessons.

\subsection{Łódź Tram Derailment (2008)}

The Łódź tram event is a reminder that cyber-physical manipulation does not always require a modern network. The reported use of a modified remote control to influence points illustrates a basic security principle: a control channel without source authentication can become operationally manipulable.

\subsection{San Francisco Muni Ransomware (2016)}

The San Francisco Muni ransomware incident demonstrates that enterprise IT compromise can disrupt service even without reaching train-control systems. Ticketing, scheduling, email, and customer-facing systems can impose operating cost, public disruption, and recovery burden while safety-critical OT remains isolated.

\subsection{Iran Railways Wiper — "MeteorExpress" (2021)}

The Iran railways wiper incident shows the sabotage potential of destructive malware. Passenger information, ministry systems, and operational coordination can be disrupted through IT-side compromise, even where there is no confirmed direct manipulation of train control.

\subsection{Belarus "Cyber-Partisans" Railway Attack (2022)}

The Belarus railway attack illustrates the convergence of hacktivist objectives and criminal tooling. Even limited disruption to ticketing or automated support functions can become strategically meaningful when railways are part of military or humanitarian logistics.

\subsection{Denmark DSB Stoppage via the Supeo Supplier (2022)}

The Denmark DSB stoppage shows how supplier dependency can become an availability risk. If drivers or dispatchers depend on a third-party application for operationally necessary information, a supplier shutdown or compromise can halt service even when the operator's own network is not breached.

\subsection{Poland Radio-Stop Abuse (2023)}

In August 2023, Polish trains were halted by abuse of an unauthenticated railway radio-stop command \cite{ReutersPoland2023}. The case is central because it reached OT directly and required neither sophisticated malware nor an IT-to-OT pivot.

\subsection{Cross-Cutting Synthesis}

\begin{table}
  \caption{Documented rail cyber incidents and their lessons.}
  \label{tab:rail_incidents}
    \renewcommand{\arraystretch}{1.2}
    \begin{tabular}{|
    >{\raggedright\arraybackslash}p{1.6cm}| 
    >{\raggedright\arraybackslash}p{2.5cm}| 
    >{\raggedright\arraybackslash}p{2.5cm}| 
    >{\raggedright\arraybackslash}p{2.5cm}| 
    >{\raggedright\arraybackslash}p{1.7cm} |
    >{\raggedright\arraybackslash}p{4.0cm}|}
      \hline
      \textbf{Incident (Year)} &
      \textbf{Actor Type} &
      \textbf{Attack Class} &
      \textbf{System Reached} &
      \textbf{Control Channel Reached?} &
      \textbf{Core Lesson} \\
      \hline

      Łódź tram (2008) &
      Individual / insider &
      Unauthenticated infrared (IR) control of points &
      Trackside points &
      Yes &
      Legacy control signals lacking source authentication can directly manipulate railway infrastructure. \\ \hline

      SF Muni (2016) &
      Cybercriminal (opportunistic) &
      Ransomware (HDDCryptor/Mamba) &
      IT systems: ticketing and scheduling &
      No &
      IT ransomware can significantly disrupt operations, while backups and network segmentation help contain the impact. \\ \hline

      Iran railways (2021) &
      State-aligned (unattributed) &
      Wiper malware (MeteorExpress) &
      IT systems and passenger information displays &
      No &
      Destructive malware can paralyze railway operations through IT systems even without compromising train-control systems. \\ \hline

      Belarus railways (2022) &
      Hacktivist (Cyber-Partisans) &
      Politically motivated ransomware &
      IT systems: e-ticketing and operational automation &
      No (capability claimed) &
      Criminal ransomware techniques are increasingly repurposed for politically motivated disruption. \\ \hline

      Denmark DSB (2022) &
      Cybercriminal (unattributed) &
      Supplier compromise leading to service shutdown &
      Supplier IT systems and driver application &
      No (indirect) &
      Third-party suppliers represent a critical attack surface capable of disrupting railway operations. \\ \hline

      Poland railways (2023) &
      Hacktivist / state-aligned &
      Unauthenticated radio-stop (RF) attack &
      Operational technology (OT): railway radio &
      Yes &
      Exploitation of insecure legacy radio protocols enables low-cost attacks that can halt train operations directly. \\ 

      \hline
    \end{tabular}
\end{table}

Three patterns emerge. First, incidents cluster around direct protocol weakness, IT-side disruption, and supplier cascade. Second, direct compromise of the safety-critical core remains comparatively rare in the public record, but legacy control channels and growing IT/OT convergence weaken that reassurance. Third, severity is not proportional to sophistication: low-cost radio abuse or supplier shutdown can still stop trains. These patterns motivate the impact assessment that follows.

\section{Impact Assessment}

The practical significance of these vulnerabilities lies in their consequences for safety, operations, economics, and critical-infrastructure continuity.

\subsection{Safety Consequences: Collision and Derailment Potential versus Fail-Safe Behaviour}

The gravest concern is direct physical harm through false movement authority, defeated interlocking, delayed warning, or manipulated train location. In many cases, fail-safe behavior converts cyber disturbance into a conservative stop rather than a collision. That is valuable, but incomplete: adversarial inputs may be plausible, coordinated, or correlated in ways that traditional safety cases did not assume \cite{Chen2015,Collins2023}.

\subsection{Operational and Economic Disruption}

Even where safety holds, availability loss can be severe. A radio command may stop trains; ransomware may disable ticketing, scheduling, or dispatch support; and supplier outage may halt operations. Because rail underpins freight logistics and supply chains, disruption can propagate beyond the operator into production schedules, commodity flows, and public mobility \cite{FRAFreight2025}.

\subsection{Critical-Infrastructure and National-Security Dimension}

Railways are critical civilian infrastructure and, in crises, instruments of national logistics. Incidents in Belarus, Poland, and Ukraine show why rail cyber resilience belongs in transportation policy, emergency management, and national-security planning \cite{ReutersUkraine2025a,ReutersUkraine2025b,TSA2026}.

\subsection{The Global South and Emerging-Economy Dimension}

Emerging-economy railways face a specific modernization challenge: imported signalling, rolling-stock, communication, and cloud stacks may increase dependency on a small number of suppliers while local cyber workforce capacity remains limited. Less-digitized networks may retain manual fallback, but modernization can erode that resilience unless degraded operation, knowledge transfer, and local assurance capability are planned from procurement onward.

\section{Countermeasures and Defensive Architecture}

No single control secures an automated railway. Resilience requires layered technical controls, lifecycle assurance, practical degraded operation, and trained personnel who can recognize and respond when automation becomes untrustworthy.

\subsection{Security-by-Design and Defense-in-Depth: Segmentation, Zones, and Conduits}

\begin{figure}
    \includegraphics[width=0.8\linewidth]{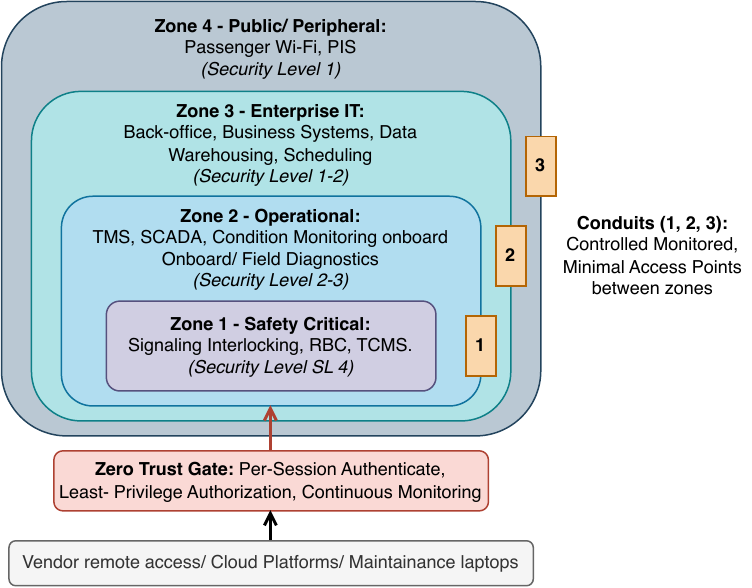}
    \caption{Zone-based zero-trust architecture for railway cyber-physical systems.}
    \label{fig:figure6}
\end{figure}

Security should be designed into procurement, architecture, testing, and change management. Defense-in-depth separates safety-critical control, maintenance, enterprise IT, passenger information, analytics, and vendor-access environments into zones connected only through monitored conduits. Figure \ref{fig:figure6} illustrates this approach. Segmentation reduces the chance that an enterprise compromise becomes an OT event while still allowing necessary operational data exchange \cite{NISTOT2023,Prochazka2019}.

\subsection{Cryptographic Protection and the FRMCS Security Model}

Protocol weaknesses argue for authenticated and integrity-protected communications, with encryption where appropriate. The constraint is operational: controls that add latency, complexity, or maintenance burden can degrade safety-critical performance if poorly implemented. FRA PTC communications research therefore treats confidentiality, integrity, availability, and operational feasibility together rather than as independent objectives \cite{Collins2023,FRAPTCComms2025}.

\subsection{OT Anomaly Detection and Zero-Trust Maintenance Access}

Prevention must be complemented by OT-aware monitoring. Detection should look for inconsistent physical data, abnormal device behavior, unauthorized configuration changes, unusual remote access, and disagreement among independent sources \cite{KourDefender2020,KourPredictive2020}. Vendor sessions, maintenance laptops, and privileged access should be authenticated per session, limited by least privilege, routed through controlled jump hosts, and logged continuously \cite{NISTOT2023}.

\subsection{Resilience and Safe-State Fallback}

Because prevention and detection will sometimes fail, automated railways need defined degraded modes: backup communications, local control, restricted movement, manual inspection, temporary crossing protection, and restoration criteria. Recovery must reestablish trusted software, configuration, credentials, data, and physical conditions before normal automation resumes. Meaningful human oversight is the adaptive layer that makes these fallbacks practical \cite{FrancePeterson2023,Thron2024,TSA2026}.

\section{Standards, Governance, and Regulation}

Rail cybersecurity is increasingly governed by standards and regulations that connect security to the established railway safety lifecycle.

\subsection{The CENELEC Safety Triad and Its Cybersecurity Extension}

Railway safety engineering rests on lifecycle standards for reliability, availability, maintainability, safety, software, signalling, and safety-related communication. Historically, these standards emphasized accidental failure. Automated, connected railways require cybersecurity to be integrated into the same assurance lifecycle, because a cyber event can create or conceal a hazardous state.

\subsection{CLC/TS 50701 and IEC 62443}

CLC/TS 50701 adapts industrial-control cybersecurity concepts to railway applications and aligns them with the rail safety lifecycle. IEC 62443 contributes the zones-and-conduits model, security levels, and secure product-development concepts. Table \ref{tab:rail_standards} maps the principal instruments and their roles.

\begin{table}[!ht]
  \caption{The rail cybersecurity standards and governance landscape.}
  \label{tab:rail_standards}
    \renewcommand{\arraystretch}{1.2}
    \begin{tabular}{|
    >{\raggedright\arraybackslash}p{4.2cm} |
    >{\raggedright\arraybackslash}p{2.8cm} |
    >{\raggedright\arraybackslash}p{8cm}|}
      \hline
      \textbf{Instrument} &
      \textbf{Type} &
      \textbf{Role} \\
      \hline

      EN 50126 / EN 50128 / EN 50129 (+ EN 50159) &
      Safety standards &
      Defines the RAMS lifecycle, software safety, signalling-system safety, and safety-related communication requirements for railway systems. \\ \hline

      IEC 62443 series &
      Cybersecurity standard &
      Provides a comprehensive OT cybersecurity framework, including zones and conduits, security levels (SL~1--4), and secure product development practices. \\ \hline

      CLC/TS 50701:2023 &
      Rail cybersecurity specification &
      Adapts IEC~62443 to railway applications and integrates cybersecurity into the EN~50126 lifecycle through risk assessment, zoning, and assurance activities. \\ \hline

      NIST SP~800-82 Rev.~3 / SP~800-207 &
      Guidance &
      Offers guidance for industrial control system (ICS/OT) security and principles for implementing zero-trust architecture. \\ \hline

      NIS2 Directive (EU) 2022/2555 &
      Regulation &
      Establishes cybersecurity obligations for essential entities, including rail operators, covering incident reporting and supply-chain security. \\ \hline

      TSA Security Directive 1580-21-01E &
      Regulation (US) &
      Specifies mandatory cybersecurity and incident-response requirements for covered U.S. railroads. \\ \hline

      Europe's Rail Joint Undertaking (formerly Shift2Rail) &
      R\&D programme &
      Supports research and innovation for secure, digital railway systems, including integrated safety--security assurance. \\ 

      \hline
    \end{tabular}
\end{table}

\subsection{Regulatory Context: NIS2, TSA Directives, and Europe's Rail}

Regulation also compels action. In the United States, TSA Security Directive 1580-21-01E requires covered freight railroads to maintain cybersecurity implementation and incident-response capabilities \cite{TSA2026}. In Europe, rail operators are increasingly treated as critical or essential entities with obligations for risk management, incident reporting, and supply-chain security. These regimes reinforce the same operational point: cybersecurity must be managed as part of railway continuity.

\section{Discussion}

\subsection{The Safety–Security Integration Tension and Certification Friction}

The central unresolved tension is between safety certification and cybersecurity response. Safety assurance is evidence-heavy and based on stable failure modes; cybersecurity is adversarial and changes as vulnerabilities, tools, and actors evolve. A security patch may affect a safety case, while slow safety recertification may leave known vulnerabilities exposed. The practical direction is an integrated safety-security assurance case in which hazards, threats, mitigations, residual risk, human roles, and change management are evaluated together \cite{Chen2015,NISTOT2023}.

The four research questions point to the same conclusion from different angles. RQ1 is answered by the IT/OT convergence pathway: automation expands the attack surface by moving safety-relevant functions onto shared software, communication, identity, and vendor-service dependencies. RQ2 is answered by the threat-surface taxonomy: communications, train location, remote access, software updates, and supplier dependencies carry the greatest potential for cascading operational effects. RQ3 is answered less completely, which is itself an important finding: distinguishing cyber compromise from equipment fault requires independent evidence, data provenance, operational context, and OT-aware monitoring rather than network alarms alone. RQ4 is answered by the resilience framework: segmentation, authenticated communications, controlled maintenance access, protected updates, degraded operation, trusted recovery, and meaningful human oversight must work together.

\subsection{Research and Capability Gaps}

Several gaps recur. Rail-specific labeled cyber datasets are scarce, and detection systems often struggle to distinguish attack from ordinary equipment fault under operational variability \cite{QiWang2024}. The workforce gap is equally important: automated railways need personnel who understand operations, signalling, OT security, human factors, and incident recovery \cite{KourKarim2021,Thron2024}. Human-oversight design also remains under-evaluated before deployment, particularly for rare, high-consequence cyber-physical events.

This study has limits. It is a narrative synthesis, not a systematic review or statistical estimate of rail cyber incident frequency. Public incident reporting is incomplete, uneven across regions, and biased toward visible disruptions. Standards and regulatory examples also reflect the sources available for North American and European rail systems more than the full range of global practice. The value of the paper is therefore not in counting events, but in organizing the operational pathways through which automation-related cyber failures can affect rail service.

\section{Future Directions}

Four developments will shape the agenda. First, FRMCS/5G migration can improve security while introducing IP-based complexity and coexistence risk with legacy systems. Second, AI-based OT monitoring may improve anomaly detection but introduces adversarial inputs, model drift, opacity, and false-alarm burden \cite{QiWang2024,Unger2023}. Third, digital twins and testbeds can evaluate cyber-physical scenarios without endangering live operations \cite{Hosseinzadeh2024}. Fourth, long rail asset lives make cryptographic agility important because communication systems and keys may remain in service for decades.

Near-term research should move from general risk identification to operationally testable evidence. Useful work would include rail-specific incident corpora, testbed datasets that include both faults and attacks, exercises that measure degraded-mode performance, and human-in-the-loop studies of dispatcher, maintainer, and operator response during cyber-physical ambiguity. These studies would give agencies and suppliers better evidence for procurement requirements, safety cases, and training programs.

\section{Conclusion}

Rail automation improves safety, capacity, inspection, maintenance, and operational visibility, but it also moves rail from isolated electromechanical control toward networked cyber-physical dependence. The paper's threat-surface taxonomy shows that cybersecurity risk is created by more than malware. It can arise from loss of trustworthy data, communication, configuration, remote access, and supplier services. Documented incidents give a measured warning: direct compromise of the safety-critical core is still rare in the public record, yet legacy command channels, IT-side disruption, and supplier cascades already stop rail service. The appropriate response is to engineer automation as a resilient sociotechnical system: segmented architecture, authenticated communications, protected remote access, OT monitoring, practical degraded operation, trusted recovery, and human oversight designed as a system requirement.

\section{Acknowledgments}

Generative artificial intelligence tools were used to support language editing, condensation, and formatting checks. The authors reviewed and verified the manuscript content, citations, and conclusions.

\section*{AUTHOR CONTRIBUTIONS}
The authors confirm contribution to the paper as follows: 
study conception and design: Syed Owais Athar, Mohsin Kamal, Aemal Khattak, and Muhammad Naveed Aman; 
source review and synthesis: Syed Owais Athar and Muhammad Naveed Aman; 
analysis and interpretation of results: Syed Owais Athar, Mohsin Kamal, Aemal Khattak, and Muhammad Naveed Aman; 
draft manuscript preparation: Syed Owais Athar. 
All authors reviewed the results and approved the final version of the manuscript.

\section*{DECLARATION OF CONFLICTING INTERESTS}
The authors declared no potential conflicts of interest with respect to the research, authorship, and/or publication of this article.

\section*{FUNDING}
The authors disclosed no financial support for the research, authorship, and/or publication of this article.

\newpage
\bibliographystyle{trb}
\bibliography{References}
\end{document}